\documentclass{article}
\usepackage{graphicx} 
\usepackage{amsmath, amssymb} 
\usepackage{hyperref} 
\usepackage{natbib} 
\usepackage{geometry} 
\usepackage{float}
\usepackage{subcaption}
\title{Application of MATEC (Multi-AI Agent Team Care) Framework in Sepsis Care}
\author{
    Andrew Cho$^1$, Jason M. Woo$^2$, Brian Shi$^1$, Aishwaryaa Udeshi$^1$, \\ Jonathan S. H. Woo$^{*3}$, MD \\
    \small $^1$Princeton University, Princeton, NJ \\
    \small $^2$University of Pittsburgh, School of Computing \& Information, PA \\
    \small $^3$Department of Medicine, Penn Medicine Princeton Health, University of Pennsylvania Health System \\
    \small $^*$ Corresponding author: jonathan.woo@pennmedicine.upenn.edu
    }
\date{} 

\begin{document}
\maketitle

\begin{abstract}
Under-resourced or rural hospitals have limited access to medical specialists and healthcare professionals, which can negatively impact patient outcomes in sepsis. To address this gap, we developed the MATEC (Multi-AI Agent Team Care) framework, which integrates a team of specialized AI agents for sepsis care. The sepsis AI agent team includes five doctor agents, four health professional agents, and a risk prediction model agent, with an additional 33 doctor agents available for consultations. Ten attending physicians at a teaching hospital evaluated this framework, spending approximately 40 minutes on the web-based MATEC application and participating in the 5-point Likert scale survey (rated from 1-unfavorable to 5-favorable). The physicians found the MATEC framework very useful (Median=4, P=0.01), and very accurate (Median=4, P$<$0.01). This pilot study demonstrates that a Multi-AI Agent Team Care framework (MATEC) can potentially be useful in assisting medical professionals, particularly in under-resourced hospital settings.
\end{abstract}

\section{Introduction}
Studies showed team-based care improves patient outcomes and patient satisfaction.(\cite{reiss-brennan_association_2016, will_team-based_2019})  Medical care for hospitalized patients, particularly those with life-threatening conditions like sepsis, requires early, coordinated care from a multidisciplinary team of healthcare professionals.(\cite{rudd_global_2018, evans_surviving_2021}) However, under-resourced hospitals often face limited access to medical specialists and professionals, which can negatively impact sepsis outcomes. (\cite{jindal_eliminating_2023}) \vspace{0.2cm}

LLM (large language model) applications via multi-agent conversations have been used in question answering, research and coding.(\cite{swanson_virtual_2024, wu_autogen_2023}) Fan et al.\ described AI hospital which included agents representing a patient, examiner and chief physician; (\cite{fan_ai_2024}) however, real hospital settings involve doctors from many specialities and healthcare professionals such as nurses, pharmacists, social workers, case managers, and patient safety officers to meet the complex medical needs of patients.  \vspace{0.2cm}

Even though LLM-based medical diagnosis has improved over time in accuracy,(\cite{tu_towards_2024, osullivan_towards_2024})  hallucination, reliability and trust remain as barriers to medical usage.(\cite{lee_benefits_2023, zuchowski_trust_2024})  To address this limited access to doctors and current model limitations, we developed the MATEC (Multi-AI Agent Team Care) framework that integrates a team of specialized AI agents into sepsis care workflows to support diagnosis, treatment planning, risk prediction, and patient care.  This proof of concept pilot study aims to demonstrate novel use cases for implementing a team of AI agents to enhance patient safety and improve patient outcomes.(\cite{fan_ai_2024, pierson_using_2025, boussina_large_2024})

\section{Methods}
Our specialized AI agents were developed using a base language model, prompt engineering, and techniques like Chain of Thought (CoT) Reasoning, ReAct, and RAG (Retriever-Augmented Generation) as illustrated by Figure 1.(\cite{wei_chain--thought_2023, yao_react_2023, ng_rag_2025})  Chroma was used to create a vector database for RAG.(\cite{noauthor_introduction_nodate})

In the MATEC framework seen in Figure 2, AI agents were designed to work collaboratively to diagnose sepsis and develop treatment plans. The Emergency Medicine Doctor, Hospitalist, Infectious Disease Doctor, and Critical Care AI agents generate differential diagnoses, identify the most likely diagnosis through diagnostic reasoning, and propose treatment plans. The Senior Physician AI agent synthesizes the inputs from doctor AI agents, verifies facts, screens for potential hallucinations, and then provides final diagnoses and treatment recommendations. The Nurse agent generates nursing recommendations, and the Pharmacy agent provides information on medication dosage, and adverse drug effects. Agents, including a Social Worker, Patient Safety Officer and QI (Quality Improvement), address social determinants of health, quality metrics like SEP-1 compliance, and hospital-acquired complication prevention. Meanwhile, a Risk Prediction Model AI agent assesses the risks of patient deterioration using tools such as the National Early Warning Score (NEWS).(\cite{holland_united_2023, covino_predicting_2023}) \vspace{0.2cm}

To evaluate the MATEC framework’s usefulness, accuracy, and relevance, we conducted a survey with ten attending physicians (hospitalist and internal medicine physicians) following a 40-minute session interacting with our MATEC framework on a web interface (Figure 3). These physicians evaluated the framework using six questionnaires on a 5-point Likert Scale (1 = unfavorable, 5 = favorable). Evaluation cases included those from the New England Journal of Medicine cases and detailed hypothetical clinical vignettes involving various sepsis scenarios. 

A one-sample Wilcoxon signed-rank test was used to evaluate whether the observed median responses significantly deviated from a neutral score of 3. R statistical software (RStudio 1.4.1717) was used for statistical analysis. \vspace{0.5cm}

\begin{figure}[H]
    \centering
    \includegraphics[width=0.95\linewidth]{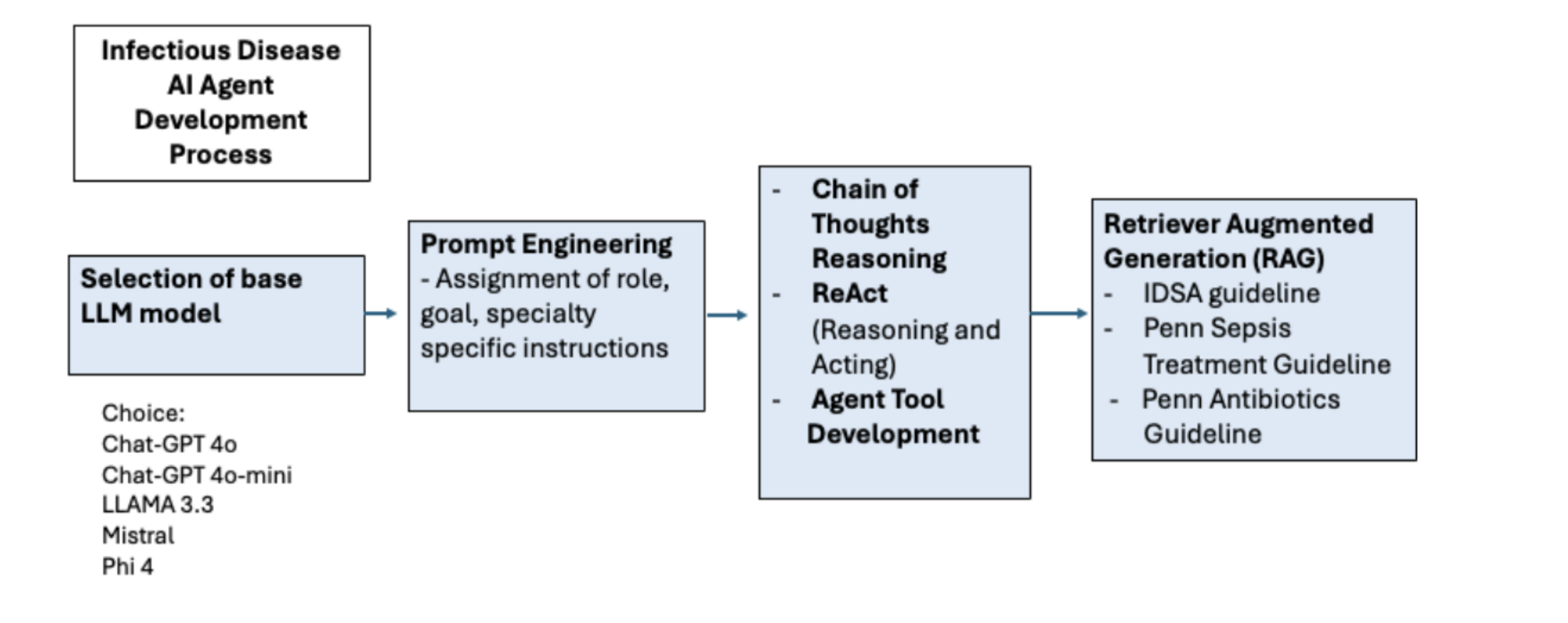}
    \caption{}
    \label{fig:fig1}
\end{figure}

\begin{figure}[H]
    \centering
    \includegraphics[width=1\linewidth]{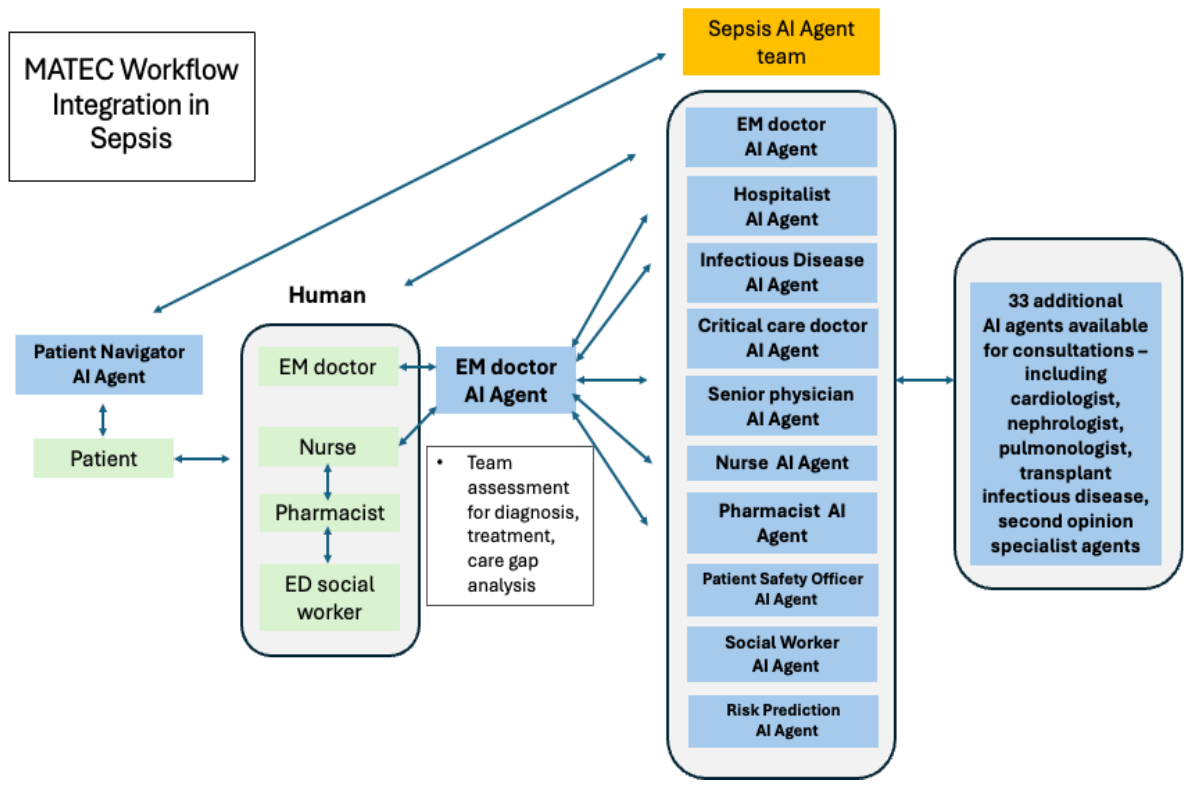}
    \caption{}
    \label{fig:fig2}
\end{figure}

\begin{figure}[H]
    \centering
    \includegraphics[width=1\linewidth]{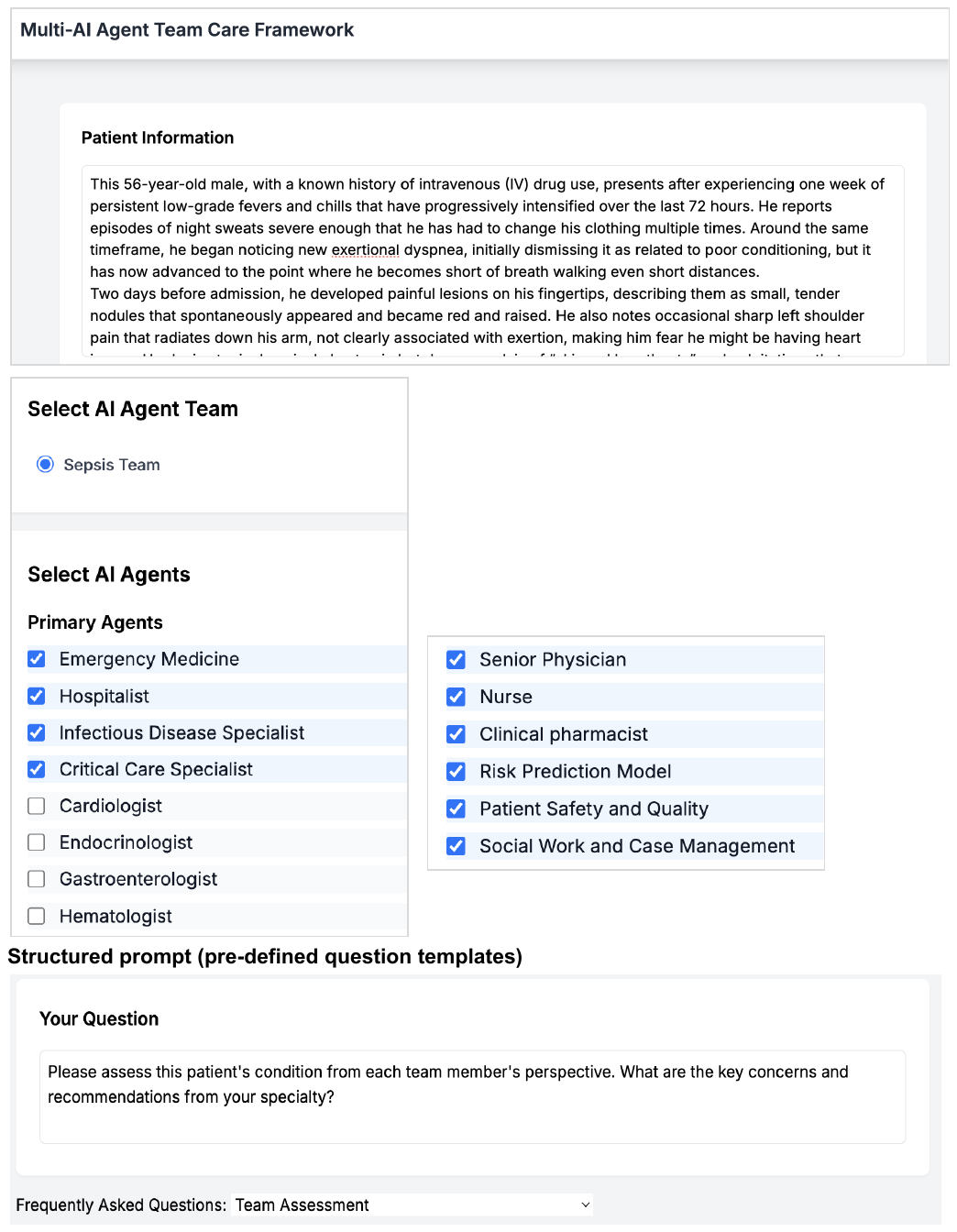}
    \caption{MATEC framework web-based user interface.}
    \label{fig:fig3}
\end{figure}

\section{Results}
A user submits patient information and a question using a user interface (Figure 3). Each attending physician selected various clinical scenarios and cases to evaluate and test the framework. They evaluated the accuracy, usefulness, and consistency of the sepsis team agents’ responses, as well as the performance of randomly selected agents among 33 other agents.  \vspace{0.5cm}

\noindent\textbf{Framework Usefulness} 

10 attending physicians rated the framework to be very useful (Median = 4, P = 0.01). Four physicians selected 5 (highly useful), and four chose 4 (very useful), as seen in Figure 6. 

Among the ten physicians surveyed, six had prior experience using LLMs like ChatGPT, and they rated this framework as more useful compared to other LLMs (Median = 4.5, P = 0.03). Users found structured prompts (pre-defined question templates) under Frequently Asked Questions very useful (Median = 5, P = 0.005).  Examples of structured prompts are shown in table 1.  \vspace{0.5cm}

\noindent\textbf{Framework Accuracy} 

The participants rated the framework as very accurate (Median = 4, P = 0.005) after testing the framework on various clinical scenarios. Six physicians rated the framework as very accurate (4) and four physicians rated it as highly accurate (5), as illustrated in Figure 6. The example of the senior doctor agent output is shown in Figure 4. The senior physician agent summarizes the final diagnosis, areas of consensus/divergence, comprehensive care plan, and next steps based on inputs of four other doctor agents in the sepsis care team. Users also rated the consistency of AI agent responses as very consistent (Median = 4, P = 0.007).\vspace{0.5cm}

\noindent\textbf{Care Gap Analysis}

The sepsis team agents were designed to perform care gap analysis. Users can select a predefined question template for this purpose, such as:  “Are there any gaps in the current care plan? Please identify potential areas for improvement in diagnosis, treatment, monitoring, and care coordination from your specialty perspective.”

Each agent evaluates human doctors’ and professionals’ evaluations, diagnoses, and treatments, identifying potential areas for improvements based on its specialty.  After individual agent assessment, users receive a summarized analysis from the sepsis team. An example of care gap analysis is shown in Figure 5. Study participants rated this as very useful (Median = 4, P = 0.02). In the provided example, the analysis highlighted suboptimal treatment of substance abuse disorder, delays in early intervention for endocarditis diagnosis, and the need for social services to address homelessness.\vspace{0.5cm}

\noindent\textbf{Social Determinants of Health}

Social determinants of health (SDOH) have been shown to significantly influence health outcomes.(\cite{hood_county_2016}) However, the collection and utilization of SDOH data for both individual and population health management remain unreliable.(\cite{truong_utilization_2020}) 

To effectively collect SDOH information and incorporate it into treatment planning,(\cite{guevara_large_2024}) a social worker agent was assigned to assess patients for SDOH and identify potential barriers to care. This ensures that comprehensive care planning incorporates these vital information. An example of social worker agent output is shown in Figure 4. 

\begin{figure}[H]
    \centering
    \includegraphics[width=1\linewidth]{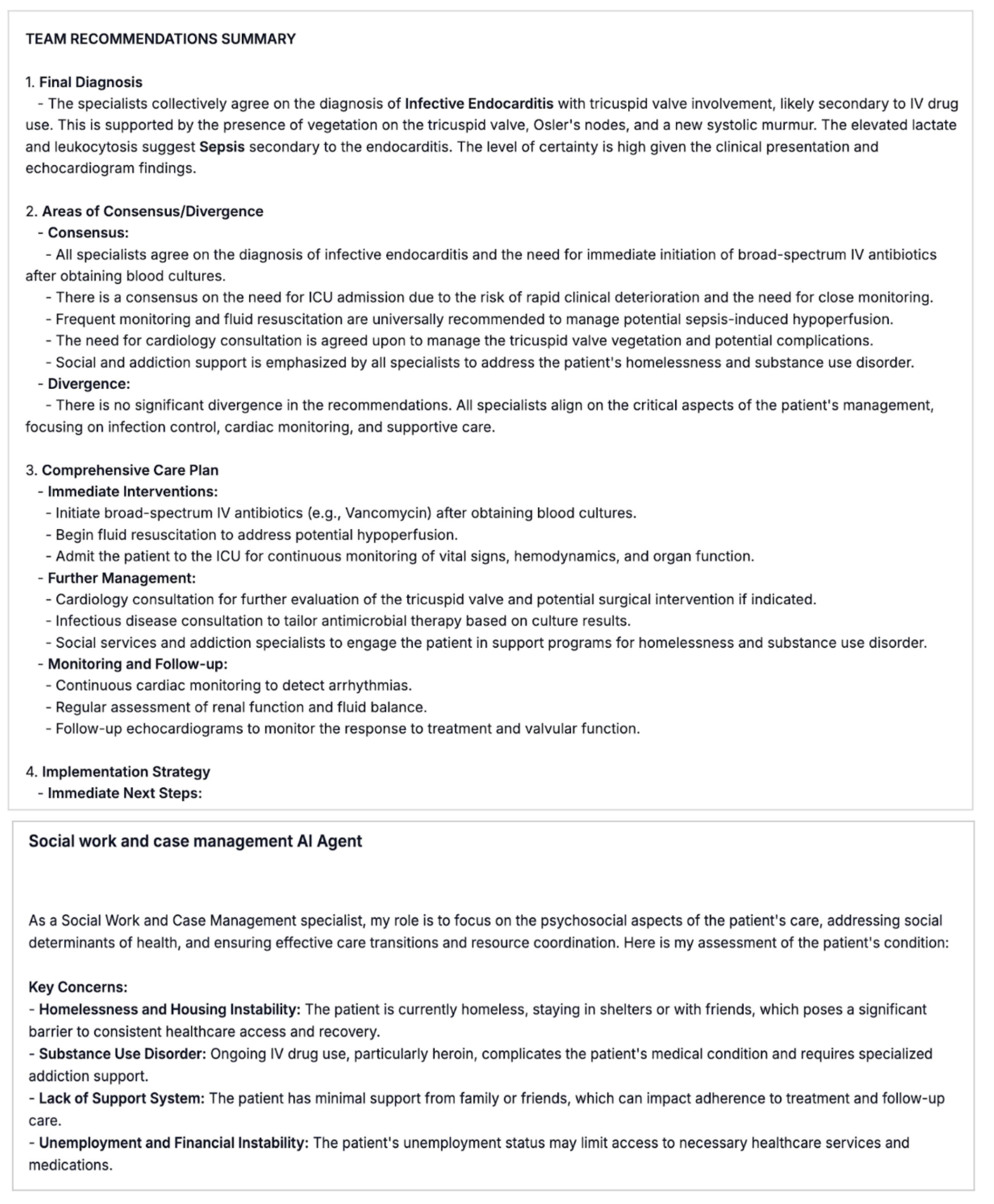}
    \caption{Example output from sepsis medical team and social work AI agent. A case with sepsis due to endocarditis}
    \label{fig:fig4}
\end{figure}

\begin{table}[H]
    \centering
    \begin{tabular}{| p{5cm} | p{10cm} |}
        \hline
        \rule{0pt}{12pt} Team Assessment & Assess the patient’s condition from each team member’s perspective. Identify key concerns and recommendations from your specialty. \rule[-6pt]{0pt}{6pt} \\
        \hline
        \rule{0pt}{12pt} Care Gap Analysis & Identify gaps in the current care plan, including potential improvements in diagnosis, treatment, monitoring, and care coordination. \rule[-6pt]{0pt}{6pt} \\
        \hline
        \rule{0pt}{12pt} Differential Diagnosis Analysis & Provide a differential diagnosis based on the patient's presentation and clinical findings, with reasoning and supporting evidence. \rule[-6pt]{0pt}{6pt} \\
        \hline
        \rule{0pt}{12pt} Treatment Plan & Recommend a treatment plan, including immediate interventions and long-term management strategies. \rule[-6pt]{0pt}{6pt} \\
        \hline
        \rule{0pt}{12pt} Antibiotic Management & Determine the appropriate antibiotic regimen, considering local resistance patterns, patient factors, and current guidelines. \rule[-6pt]{0pt}{6pt} \\
        \hline
        \rule{0pt}{12pt} Pharmacy Assessment & Assess medication management and pharmaceutical care considerations, including medication safety and monitoring. \rule[-6pt]{0pt}{6pt} \\
        \hline
    \end{tabular}
    \caption{Structured prompt examples (pre-defined question templates) under Frequently Asked Questions}
    \label{tab:patient_assessment}
\end{table}

\begin{figure}[H]
    \centering
    \includegraphics[width=1\linewidth]{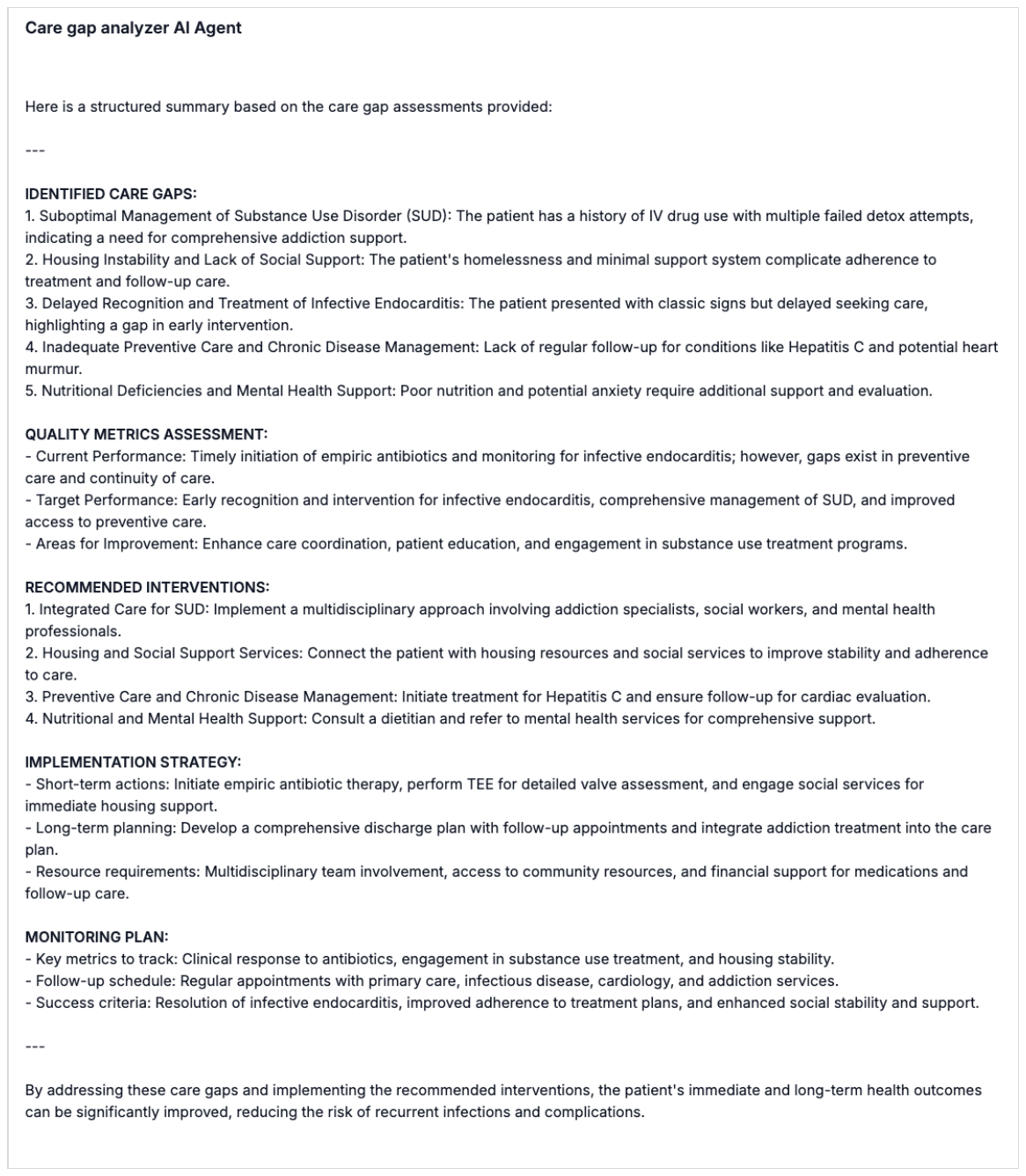}
    \caption{Example of Care Gap Analysis.}
    \label{fig:fig5}
\end{figure}

\begin{figure}[H]
    \centering
    \includegraphics[width=1\linewidth]{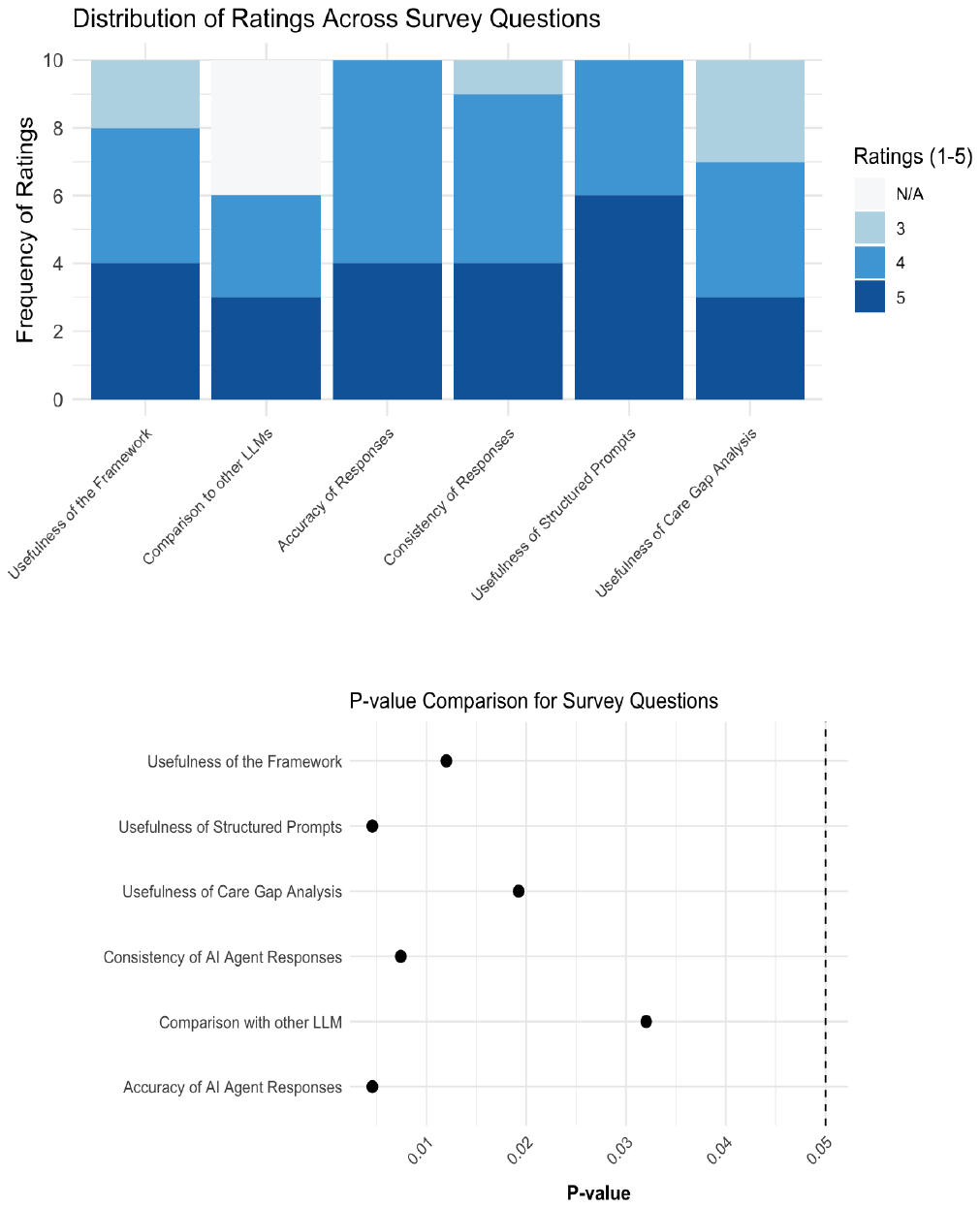}
    \caption{Stacked bar chart illustrating distribution of survey results based on Likert-scale from 1 to 5.}
    \label{fig:fig6}
\end{figure}

\section{Discussion}
The MATEC framework was developed to address the limited access to specialists and health professionals affecting rural and under-resourced hospitals. This framework demonstrates how specialized AI agents can aid human medical teams by assisting in diagnosis, treatment planning, and the identification of care gaps.  \vspace{0.2cm} 

The collaborative of the Institute of Medicine roundtable outlined key principles of team-based healthcare, including shared goals, clearly defined roles, mutual trust, effective communication, and measurable processes and outcomes.(\cite{mitchell_core_2012}) To align with these principles, the MATEC framework incorporates system prompts to establish shared goals of the sepsis team, define clear roles among doctor and health professional agents, and ensure transparent reasoning and diagnostic processes to foster mutual trust and effective communication. In addition, the framework includes a patient safety and quality improvement (QI) agent to monitor compliance with SEP-1 (Sepsis and Septic Shock Early Management Bundle) and hospital acquired infection prevention measures. (\cite{boussina_large_2024}) Furthermore, a social determinants of health assessment was performed by a social work AI agent as studies showed the significant association of bloodstream infection rates with areas with higher poverty levels and drug use.(\cite{guevara_large_2024, quagliarello_strains_2002, rha_vital_2023}) \vspace{0.2cm}

Study participants rated the framework favorably (Figure 6). Ten attending physicians found the MATEC framework very useful (Median=4, P=0.01) and very accurate (Median=4, P$<$0.01). The high accuracy rating of this framework indicates that multiple agents cross-verifying outputs can effectively minimize errors and hallucinations. Participants also rated this framework as more useful than other LLMs.    \vspace{0.2cm}

There are multiple ways that the framework can be potentially applied in clinical settings. When a patient presents to an emergency department with initial complaints, the framework can assist the human team via decision support: the MATEC agents can assist with initial diagnosis, diagnostic work-ups, and treatment planning. The emergency medicine doctor agent receives a summary of the sepsis team agents’ comprehensive assessment and care gap analysis, identifying areas that may be inadequately addressed from medical, patient safety, quality, and social work perspectives. 

After hospital admission, the MATEC sepsis team can monitor treatment responses and suggest potential treatment modifications to the human team. The risk prediction model agent monitors patient condition at predetermined intervals and provides recommendations when there are signs of deterioration. The pharmacist agent will monitor medication dosages, drug interactions, and potential side effects. Users can also consult any specialist agent if there are specific needs. For example, while a cardiologist agent is not a part of the core sepsis team, it can be consulted for endocarditis if suggested by the sepsis agent team. 

During the pilot study, participants had access to 33 specialist agents–including a nephrologist, pulmonologist, and transplant infectious disease agent–beyond the 10 core sepsis team agents. The framework is designed for future scalability, allowing the deployment of additional specialist agents as needed.  A patient navigator agent shown in Figure 2 can explain medical teams’ treatments to a patient in understandable language and help patients better understand their illnesses (Figure 7). When a patient is ready for discharge, a case management agent can assist the human team with identifying barriers to discharge and needs for home health services. The sepsis team can provide the summary of hospitalization to the primary care doctor or agent.(\cite{hartman_developing_2024, zaretsky_generative_2024})  Care gap analysis content can also be customized depending on who is using it. A summary of care gap analysis of all patients in a particular hospital unit can be sent regularly to a unit medical director, nurse manager and safety officer to optimize care and reduce preventable complications. The MATEC framework was developed to address the variety of tasks mentioned above, but it requires further development to ensure reliable and consistent performance. Even though this pilot study was done using a web-based application for a proof of concept, the application will need to be embedded into electronic health records to better integrate with existing workflows. \vspace{0.2cm}

The MATEC framework has multiple advantages compared with a single LLM-based approach. First, multiple agents can reduce the cognitive burden on human doctors by assisting with diagnoses and treatment planning after comprehensive assessments with multiple specialists’ perspectives.  Second, its modular structure makes it less prone to model drift and deterioration by enabling the use of different base LLM models.(\cite{barabucci_combining_2024, touvron_llama_2023}) The LLM model’s performance significantly depends on its training data. As newer LLM models increasingly use synthetic data for model training and techniques like distillation, newer models need to be tested for their accuracy and reliability prior to wide adoption.(\cite{li_synthetic_2023, smolyak_large_2024, liu_large_2024}) Third, the framework’s multiple agents review each others’ outputs for fact verification to reduce hallucinations, unlike a single LLM-based approach. Furthermore, the structured approach of the sepsis agent team ensures the consistent inclusion of safety measures and social determinants of health assessment with dedicated patient safety and social work agents instead of focusing solely on diagnosis and immediate treatment. 

The MATEC framework also has limitations. Although this pilot study showed our multi-AI agent team-care framework has excellent accuracy and usefulness rated by evaluators, the technologies behind AI agents are in their early stage and rapidly advancing. More advanced agent technologies will enhance the capabilities and efficacy of the MATEC framework.  

\section{Conclusion}
This proof of concept pilot study demonstrates that the Multi-AI Agent Team Care (MATEC) framework can potentially be used to assist medical professionals, particularly in under-resourced hospital settings. More studies are necessary for applications of the multi-agent framework in clinical settings. 

\bibliographystyle{plainnat}
\bibliography{zreferences}

\appendix
\section{Supplement Figure 7}

\begin{figure}[h]
    \centering
    \includegraphics[width=1\linewidth]{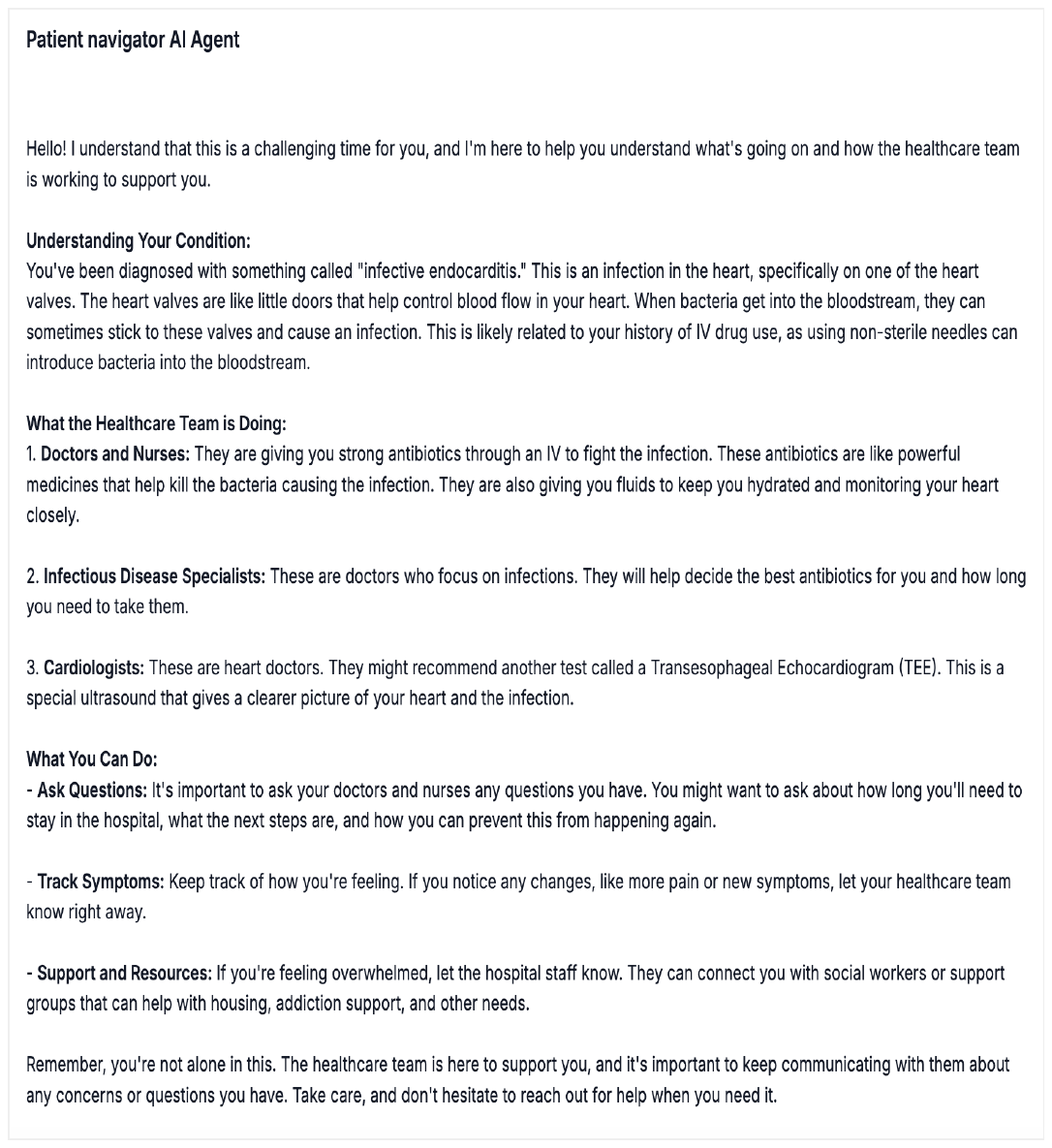}
    \caption{Example output of Patient Navigator agent.}
    \label{fig:fig7}
\end{figure}
\end{document}